\documentclass
[aps,
pre,
twocolumn,
singlespace,
groupedaddress]
{revtex4-1}

\usepackage{amsthm}
\usepackage{subfigure}
\usepackage{graphicx}
\usepackage{epsfig}
\usepackage{epstopdf}
\usepackage{color}
\usepackage{natbib}
\usepackage{amsmath}
\usepackage{amssymb}
\usepackage{hyperref}

\begin{document}

\title{Stability of miscible Rayleigh-Taylor fingers in porous media with non-monotonic density profiles}


\author{Satyajit Pramanik}
\email[]{satyajit.math16@gmail.com}
\affiliation{Nordita, Royal Institute of Technology and Stockholm University, Stockholm, Sweden}
\affiliation{Department of Mathematics, Indian Institute of Technology Ropar, Punjab, India}

\author{Manoranjan Mishra}
\email[]{manoranjan.mishra@gmail.com}
\affiliation{Department of Mathematics, Indian Institute of Technology Ropar, Punjab, India}
\affiliation{Department of Chemical Engineering, Indian Institute of Technology Ropar, Punjab, India}


\date{\today}

\begin{abstract}
We study miscible Rayleigh-Taylor (RT) fingering instability in two-dimensional homogeneous porous media, in which the fluid density varies non-monotonically as a function of the solute concentration such that the maximum density lies in the interior of the domain, thus creating a stable and an unstable density gradients that evolve in time. With the help of linear stability analysis (LSA) as well as nonlinear simulations the effects of density gradients on the stability of RT fingers are investigated. As diffusion relaxes the concentration gradient, a non-monotonic density profile emerges in time. Our simple mathematical treatment addresses the importance of density gradients on the stability of miscible Rayleigh-Taylor fingering. In this process we identify that RT fingering instabilities are better understood combining Rayleigh number (Ra) with the density gradients--hence defining a new dimensionless group, gradient Rayleigh number (Ra$_g$). We show that controlling the stable and unstable density gradients, miscible Rayleigh-Taylor fingers are controllable. 
\end{abstract}



\maketitle

\section{Introduction \label{sec:Intro}}
There are several situations in which hydrodynamic instabilities occur due to an adverse density difference in the gravity field when a heavy fluid overlies a light fluid. The study of Rayleigh-Taylor (RT) fingering in immiscible fluids dates back to the seminal work by Rayleigh \cite{Rayleigh1882}. In RT, a convective motion generates at the interface between two fluids. Understanding RT fingering is important in various geological and industrial applications such as carbon-dioxide (CO$_2$) sequestration, planetary dynamics and formation of stars \cite[][and refs. therein]{Gopal2017}. The possibility to store anthropogenic CO$_2$ in underground reservoir has renewed the interest to understand miscible RT fingers in porous media \cite{Huppert2014, Slim2014, Nakanishi2016, Binda2017}. 

Notwithstanding the intensive theoretical, numerical and experimental researches of this classical hydrodynamic instability in porous media, several questioned remained unanswered. Recently, Gopalakrishnan \emph{et al.} \cite{Gopal2017} have experimentally and numerically verified the relative role of convective and diffusive mixing in the miscible RT fingering in porous media. Through systematic analysis these authors concluded that in the fingered zone both diffusion and convection have equal contributions. The other important aspect that has not been understood yet is the role of the ``transient density gradients" on the miscible RT fingering. In miscible fluids, relaxation of the concentration gradient changes the density gradient within the diffusive layer, unlike its immiscible counterparts. 

Despite the fact that the density gradient continuously changes, viscous RT fingering in miscible fluids in a porous medium is generally characterized in terms of the Rayleigh-Darcy number \cite{Riaz2006}, 

\begin{equation}
\label{eq:Rayleigh}
{\rm Ra} = \frac{ \kappa g l_c \Delta \rho }{ \phi \mu D }, 
\end{equation}
where $\Delta \rho$ is the density difference between the two fluids, $\kappa$ is the permeability of the medium, $\mu$ is the dynamic viscosity of the fluid, $D$ is the diffusivity of the solute, $l_c$ is a characteristic length scale, $\phi$ is the fluid volume fraction (porosity), and $g$ is the gravitational acceleration. For a steady-state solution of the base solute profile, the critical Rayleigh-Darcy number for the occurrence of the instability is ${\rm Ra}_{c} = 4 \pi^2$ \cite{Horton1945}, which is violated while considering a variation in the dynamic viscosity of the fluid \cite{Sabet2017b}. 

The scenario is completely different when the base-state solution of the solute profile is time dependent--the critical Rayleigh-Darcy number depends on the onset of instability $t_{c}$, i.e., ${\rm Ra}_{c}(t_c)$. On the other hand, in reactive systems a non-monotonic density profile develops in time \cite{Lemaigre2013, Loodts2014, Loodts2016}. Similar non-monotonic profile is also evident in double diffusive systems \cite{Trevelyan2011} or as an influence of non-ideal mixing. Water density maximizes at temperature $\approx 4^{\circ}$C. This results penetrative convection in thermal Rayleigh-B\'enard convection \cite[][and refs. therein]{Srikanth2017}. It is evident from equation \eqref{eq:Rayleigh} that Ra considers the density difference, not its gradient. Therefore, we define gradient Rayleigh-Darcy number  

\begin{equation}
\label{eq:Rayleigh_new} 
{\rm Ra}_{g}(t) = \frac{ \kappa g l_c^2 \left[ \partial \rho (t)/\partial x \right] }{ \phi \mu D }. 
\end{equation}
Note that for thermal Rayleigh-B\'enard convection, the temperature field controlling the fluid density has a constant gradient. In that case the partial derivative of the local density gradient becomes an ordinary derivative and we can write $\displaystyle \frac{ {\rm d} \rho }{ {\rm d} x } = \frac{ \Delta \rho }{ l_c }$ for $l_c$ equal to the domain depth, and hence, as described in equation \eqref{eq:Rayleigh_new} the gradient Rayleigh-Darcy number (Ra$_g$) equals the Rayleigh-Darcy number (Ra). 

In this paper we show that the miscible RT fingers are better explained in terms of local density gradients, which is best described through Ra$_g(t)$, not Ra. We model the fluid density as piecewise smooth functions of a solute concentration and perform a linear stability analysis based on the widely used method of quasi-steady state approximation (QSSA) \cite{Tan1986} and nonlinear simulations of the full problem. Numerical simulations for both monotonic and non-monotonic density profiles confirm linear stability predictions. The remainder of the paper is organized as follows. In Section \ref{sec:math}, we briefly discuss a general mathematical formulation for the macroscopic description of the fingering instabilities in porous media. Results from QSSA and nonlinear simulations are discussed in section \ref{sec:results}. Finally, discussion and conclusions are presented in \S \ref{sec:discuss}. 

\section{Mathematical model and stability analysis \label{sec:math}} 

Consider the fluid motion in a two-dimensional, vertical, homogeneous porous medium. The dynamics viscosity of the fluid $\mu$ is a constant, while the density ($\tilde{\rho}$) depends on a solute concentration ($C$) such that $\tilde{\rho}(C = C_0) = \tilde{\rho}_1$ and $\tilde{\rho}(C = 0) = \tilde{\rho}_2$, which correspond to the density of the penetrating and defending fluids, respectively. Furthermore, we assume that these two fluids are initially separated by a sharp interface ($C = C_0/2$) and it expands in time, as concentration diffuses. 

\subsection{Equations of motion \label{subsec:equations}} 

A general macroscopic description of single species miscible RT fingering instabilities in two dimensional porous media is given by the following coupled nonlinear partial differential equations: \cite{Tan1986, Homsy1987, Pramanik2015e} 

\begin{subequations}
\begin{align}
& \boldsymbol{ \nabla_X } \cdot \boldsymbol{U}( \boldsymbol{X} ) = 0, \label{eq:continuity_dim} \\ 
& \boldsymbol{ \nabla_X } P = -\frac{ \mu }{ \kappa } \boldsymbol{U}( \boldsymbol{X} ) + \tilde{\rho} ( \boldsymbol{X}, \tau ) \boldsymbol{g}, \label{eq:Darcy_dim} \\
& \phi \partial_{\tau} C(\boldsymbol{X}, \tau ) + \boldsymbol{ \nabla_X } \cdot \left[ \boldsymbol{U}( \boldsymbol{X} ) C( \boldsymbol{X}, \tau ) \right] \nonumber \\ 
& \;\;\;\;\;\;\;\;\;\;\;\;\;\;\;\;\; = \boldsymbol{ \nabla_X } \cdot \left[ \phi D \boldsymbol{ \nabla_X } C( \boldsymbol{X}, \tau ) \right], \label{eq:c_dim}
\end{align}
\end{subequations}
in $( \boldsymbol{X}, \tau ) \in \Omega \times (0, \infty)$ wherein $\boldsymbol{ \nabla_X } = \displaystyle \left( \frac{ \partial }{ \partial X }, \frac{ \partial }{ \partial Y } \right)$. We assume that the gravity acts in the positive $X$-direction. Here $\boldsymbol{U}$ is the two-dimensional Darcy velocity vector, $p$ is the fluid pressure, $\mu$ is the dynamic viscosity and $\tilde{\rho}$ is the density of the fluid, $\phi$ and $\kappa$ are the porosity and permeability of the medium, and $D$ is a constant dispersion coefficient of the solute $C$. 

\subsection{Boundary and initial conditions \label{subsec:BC_IC}} 

The domain of definition of our problem is $\Omega = [-H/2, H/2] \times (-\infty, \infty)$, where $H$ is the depth of the porous media. This system of equations are associated with the following boundary and initial conditions. Here, we are interested in time scales smaller than the time at which the equilibrium state is obtained for the species. This allows us to consider the cases for which the species concentration diffuses very slowly so that the diffusive fronts do not reach the boundaries. Therefore, we have the following boundary and initial conditions. At the inlet boundary, we have 

\begin{equation}
\label{eq:BC1}
\boldsymbol{U} = (0, 0), \; \partial_X C = 0, \;\;\; [X = -H/2]. 
\end{equation}
At the outlet boundary, we have 

\begin{equation}
\label{eq:BC2}
\boldsymbol{U} = (0, 0), \; \partial_X C = 0, \;\;\; [X = H/2]. 
\end{equation}
We assume periodic boundary conditions in the transverse direction. The initial condition is zero velocity and a step-like profile for the solute concentration. That is, 

\begin{subequations}
\begin{align}
& \boldsymbol{U} = (0, 0), \label{eq:IC1} \\
& C(\boldsymbol{X}, \tau = 0) = C_0(\boldsymbol{X}) =
\left\{
\begin{array}{ll}
C_0, &  X \leq 0 \\
0, & X > 0 \\
\end{array}
\right., \forall Y. 
\label{eq:IC2}
\end{align}
\end{subequations}

\subsection{Non-dimensionalization \label{sec:dimensionless}} 

This system inherits (a) a diffusive time scale, $\tau_{d} = D/U^2$, and (b) a convective time scale, $\tau_{a} = l_a/U$, associated to a convective length scale $l_a$. Here, $U = \Delta \rho \kappa g/\mu$ is the buoyancy-induced convective velocity, where $\Delta \rho = \tilde{\rho}_1 - \tilde{\rho}_2$. Evidently, diffusion occurs on a slow time scale as compared to the convection, i.e., $\tau_{d} \ll \tau_{a}$. We are interested in the stability of the diffusive front to infinitesimal perturbations. We render the above system of equations dimensionless with the following scaling \cite{Pramanik2015d, Pramanik2016a}

\begin{subequations}
\begin{align}
& \boldsymbol{u} = \frac{ \boldsymbol{U} }{ \Delta \rho \kappa g/\mu }, \; \rho = \frac{ \tilde{\rho} }{ \Delta \rho }, \; c = \frac{C}{C_0}, \label{eq:scaling1} \\ 
& \boldsymbol{x} = \frac{ \boldsymbol{X} }{ \phi D/U } = \frac{ \boldsymbol{X} }{ \phi D \mu/\Delta \rho \kappa g }, \label{eq:scaling2} \\ 
& t = \frac{ \tau }{ \phi^2 D/U^2 } = \frac{ \tau }{ \phi^2 D \mu^2/\left( \Delta \rho \kappa g \right)^2 }, \label{eq:scaling3} \\ 
& p = \frac{ P }{ \phi \mu D/\kappa } - \frac{ \tilde{\rho}_{1} }{ \Delta \rho }x, \label{eq:scaling4} 
\end{align}
\end{subequations}
from which we obtain 
\begin{subequations}
\begin{align}
& \boldsymbol{ \nabla } \cdot \boldsymbol{u} = 0, \label{eq:continuity_nondim} \\ 
& \boldsymbol{ \nabla} p = - \left[ \boldsymbol{u} - \rho(c) \boldsymbol{e}_x \right], \label{eq:Darcy_nondim} \\ 
& \partial_{t} c + \boldsymbol{u} \cdot \boldsymbol{ \nabla } c = \nabla^2 c, \label{eq:c_nondim} \\ 
& \boldsymbol{u} = (0, 0), \; \partial_x c = 0, \;\;\; [x = -{\rm Ra}/2], \label{eq:InBC_nondim} \\ 
& \boldsymbol{u} = (0, 0), \; \partial_x c = 0, \;\;\; [x = {\rm Ra}/2], \label{eq:OutBC_nondim} 
\\ 
& \boldsymbol{u} = (0, 0), \; c(\boldsymbol{x}, t = 0) = \left\{
\begin{array}{ll}
1, &  x \leq 0 \\
0, & x > 0 \\
\end{array}
\right., \forall y, \label{eq:IC_nondim} 
\end{align}
\end{subequations}
wherein $\boldsymbol{e}_x$ is the unit vector in the $x$-direction, $\boldsymbol{ \nabla } = \displaystyle \left( \frac{ \partial }{ \partial x }, \frac{ \partial }{ \partial y } \right)$, and Ra = $\displaystyle \frac{ \Delta \rho \kappa g H }{ \phi \mu D }$ is the Rayleigh number. Note that the boundary conditions in the $y$-direction are periodic. 

\subsection{Linear stability analysis \label{subsec:LSA}} 

As mentioned above, we are interested in the stability analysis of a time-dependent base flow. Following earlier studies, we assume no convection (pure diffusion of $c$) as the base flow \cite{Riaz2006, Pramanik2015d}. Subject to this condition, as described by equation \eqref{eq:c_nondim}, the mean squared displacement of the diffusive front of the species concentration grows as $4 t$ in time, $t$. For Ra $\gg \sqrt{4 t}$, the base flow can be expressed as 

\begin{subequations}
\begin{align}
& \boldsymbol{ u }_b = \boldsymbol{0}, \label{eq:base_vel} \\ 
& c_b(x, t) = \frac{ 1 }{ 2 } \text{erfc} \left( \frac{x }{ 2 \sqrt{t} } \right), \label{eq:base_conc}
\end{align}
\end{subequations}
wherein `erfc' represents the complementary error function. Diffusive characteristic length and time scales (introduced in equations \eqref{eq:scaling2} and \eqref{eq:scaling3}) allow to invoke QSSA based \cite{Tan1986} linear stability analyses (LSA) of the miscible RT fingers in porous media. 

QSSA method is classical in the stability analysis of time-dependent base flow, but in order to make our treatment reasonably self-contained we outline the main steps in the development of the key equations of the normal mode analysis. We introduce an infinitesimal perturbation to the governing equations \eqref{eq:continuity_nondim} through \eqref{eq:c_nondim} and linearized and sought solution of the following form for the perturbations in the concentration and velocity: 

\begin{equation}
\label{eq:normalmodes} 
\left[ \tilde{c}(x, y, t), \tilde{u}(x, y, t) \right] = \left[ \bar{c}(x), \bar{u}(x) \right] e^{i k y + \omega (t_0) t}, 
\end{equation}
where $\omega(t_0)$ corresponds the growth rate of the perturbations to the base state frozen at time $t_0$, and $k$ is the wavenumber of the perturbations. Eliminating the pressure ($p$) and the transverse velocity component ($v$), we obtain an eigenvalue problem \cite{Manickam1995} 

\begin{subequations}
\begin{align}
& \left( \frac{\rm d^2}{{\rm d} x^2} - \omega(t_0) - k^2 \right) \bar{c} = \frac{{\rm d} c_b(t_0)}{{\rm d} x} \bar{u}, \label{eq:LSA1} \\ 
& \left( \frac{\rm d^2}{{\rm d} x^2} - k^2 \right) \bar{u} = -k^2 \left( \frac{{\rm d} \rho}{{\rm d} c} \right)_{c_b} \bar{c}. \label{eq:LSA2} 
\end{align}
\end{subequations}
For a finite $t_0$ and $k$, the eigenvalue problem, described by Eqs. \eqref{eq:LSA1} and \eqref{eq:LSA2}, is solved numerically using a finite difference method on a uniform mesh \cite{Pramanik2013}. In section \ref{sec:results}, we present numerically computed dispersion curves $\omega(t_0) = \omega(k; t_0)$ and the eigenfunction of the most unstable modes. For a detail discussion of QSSA method, interested readers are redirected to Refs. \cite{Tan1986, Manickam1995, Pramanik2015f}. 

\subsection{Nonlinear simulations \label{subsec:NLS}} 

We use Fourier pseudo-spectral method to numerically solve the nonlinear partial differential equations \cite{Pramanik2015d}

\begin{subequations}
\begin{align}
& \nabla^2 \psi = \rho^{\prime}(c) \frac{ \partial c }{ \partial y }, \label{eq:SF1} \\ 
& \frac{ \partial c }{ \partial t } + \frac{ \partial \psi }{ \partial y } \frac{ \partial c }{ \partial x } - \frac{ \partial \psi }{ \partial x } \frac{ \partial c }{ \partial y } = \nabla^2 c, \label{eq:SF2} 
\end{align}
\end{subequations}
in a doubly periodic domain of size [Ra, Ra/$A$], where $A$ is the aspect ratio of the computational domain. Here $\psi$ is the stream function, which is defined as $ \displaystyle \boldsymbol{u} = \left( \frac{ \partial \psi }{ \partial y }, -\frac{ \partial \psi }{ \partial x } \right)$. We consider the Fourier expansion of the unknown variables and the nonlinear terms 
\begin{subequations}
\begin{align}
& M(x, y, t) = \rho^{\prime}(c) \frac{\partial c}{\partial y}, \label{eq:M} \\ 
& N(x, y, t) = \frac{ \partial \psi }{ \partial y } \frac{ \partial c }{ \partial x } - \frac{ \partial \psi }{ \partial x } \frac{ \partial c }{ \partial y }, \label{eq:N} 
\end{align}
\end{subequations}
as follows: 

\begin{subequations}
\begin{align}
& c(x, y, t) = \sum_{j, l} \hat{c}_{j, l} (t) e^{i(k_j x + k_l y)}, \label{eq:F1} \\ 
& \psi(x, y, t) = \sum_{j, l} \hat{\psi}_{j, l} (t) e^{i(k_j x + k_l y)}, \label{eq:F2} \\ 
& M(x, y, t) = \sum_{j, l} \hat{M}_{j, l} (t) e^{i(k_j x + k_l y)}, \label{eq:F3} \\ 
& N(x, y, t) = \sum_{j, l} \hat{N}_{j, l} (t) e^{i(k_j x + k_l y)}, \label{eq:F4} 
\end{align}
\end{subequations} 

In the Fourier space, the resulting differential-algebraic equations 

\begin{subequations}
\begin{align}
& \frac{{\rm d} \hat{c}_{j, l}(t)}{{\rm d} t} + \hat{N}_{j, l}(t) = -(k_j^2 + k_l^2) \hat{c}_{j, l}(t), \label{eq:DAE1} \\ 
& \hat{\psi}_{j, l}(t) = -\frac{\hat{M}_{j, l}(t)}{(k_j^2 + k_l^2)}, 
\end{align}
\end{subequations}
are solved using an operator splitting predictor-corrector method. Adams-Bashforth method is used as a predictor followed by a trapezoidal method as the corrector \cite[see][and refs. therein for the details of the numerical method]{Tan1988, Pramanik2015d, Pramanik2016a, Pramanik2015f}. 

Note that since Ra appears in the boundary conditions, the (in)stability of the system (i.e., the onset of instability, the unstable wave numbers, etc.) is independent of the choice of the value of Ra. It is verified that the numerical eigenvalues of the linear stability problem is independent for Ra $\geqslant 200$. Therefore, unless mentioned otherwise, we choose Ra = 200 in our linear stability analysis. On the other hand, for nonlinear simulations, the value of Ra imposes a cutoff time of our simulations--owing to the periodicity, simulations are terminated as soon as a finger reaches either of the two horizontal boundaries. The nonlinear simulation results shown in this paper are obtained using Ra $= 4096$ and $A = 4$. Grid independence has been verified and we use $1024 \times 256$ spectral points and $\Delta t = 0.2$ that assure an $\mathcal{O}(10^{-4})$ accuracy of our numerical results. 

\begin{figure*}[hbtp]
(a) \hspace{4.5 in} (b) \\ 
\centering 
\includegraphics[scale=0.75]{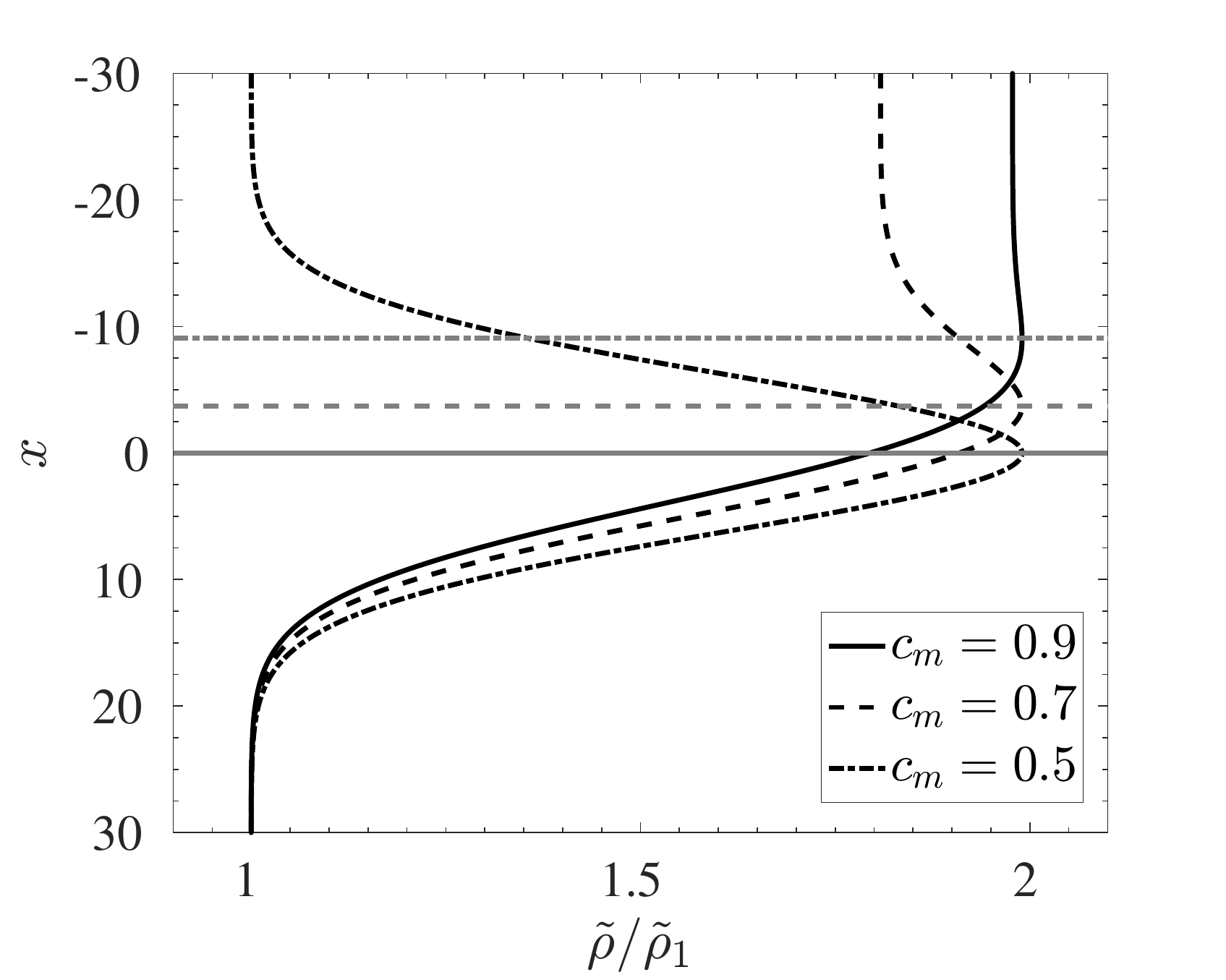} 
\includegraphics[scale=0.75]{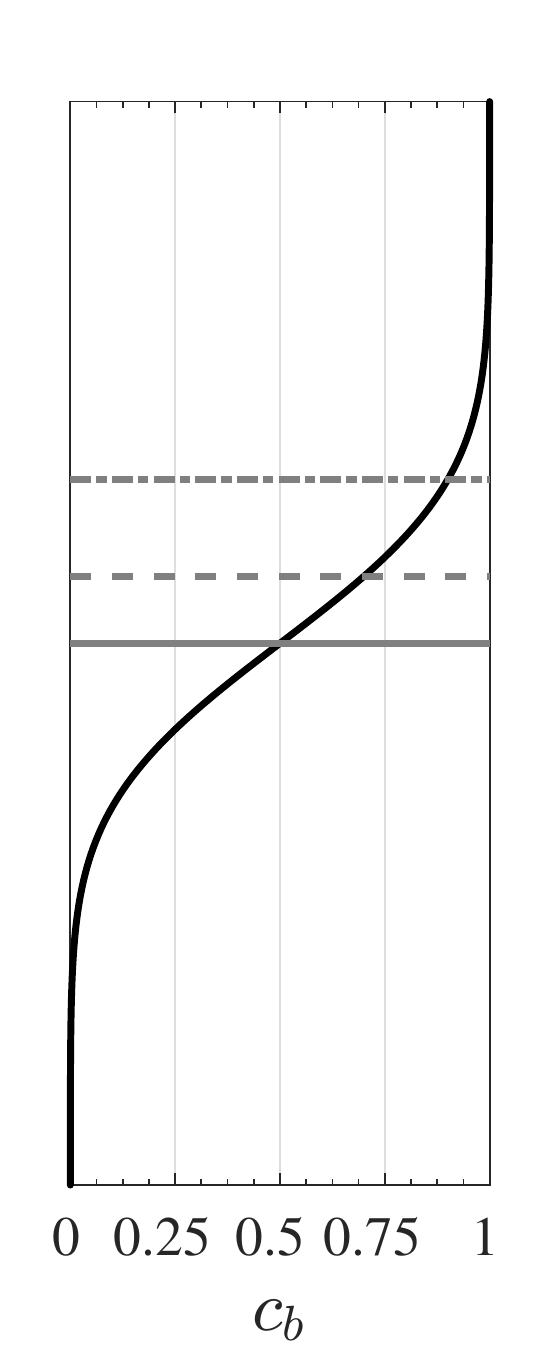}
\caption{For $\rho_m = 1.8, \; t_0 = 25$ and Ra = 200, spatial variation of (a) $\tilde{\rho}$ defined in \eqref{eq:nonmonotonic3}, for $c_m = 0.5, 0.7, 0.9$; (b) $c_b$ defined in \eqref{eq:base_conc}. The horizontal solid, dashed, and dash-dotted lines correspond to the positions of $c_b = 0.5$, 0.7, and 0.9, respectively.} 
\label{fig:nonmonotonic_densities}
\end{figure*}

\subsection{Density profile \label{subsec:density}}

To close the mathematical description of the problem, we need to define the density profile. In various reaction-diffusion and double diffusion models, a non-monotonicity in the concentration-dependent density profile emerges due to chemical reaction and differential diffusion \cite[and refs. therein]{Trevelyan2011, Lemaigre2013, Loodts2015, Loodts2016}. Chemical reaction induces asymmetry with regard to the initial contact line to an otherwise symmetric Rayleigh-Taylor or double-diffusive patterns \cite{Lemaigre2013}. We are interested to understand the basic fluid mechanical processes to understand the asymmetric finger patterns free from the ravages of such complicated situations of chemical reactions and differential diffusion. In this direction, we consider a phenomenological model for the density profile as a non-monotonic function of $c (= C/C_0)$: 

\begin{equation}
\label{eq:nonmonotonic3} 
\tilde{\rho}(c) = \tilde{\rho}_1 \left[ 1 + \frac{ 1 - \rho_m }{ c_m^2 } c (c - 2 c_m) \right], 
\end{equation}
such that, $ \displaystyle \tilde{\rho}_2 = \tilde{\rho}_1 \left[ 1 + \frac{ 1 - \rho_m }{ c_m^2 } (1 - 2 c_m) \right]$. Wherein $\displaystyle \rho_m = \left[ \max_C \{ \tilde{\rho}(C) : 0 \leq C \leq C_0 \} \right] \Biggm/ \tilde{\rho}_1$, and $c_m$ is the corresponding dimensionless value of the concentration, i.e., $\displaystyle \rho_m = \frac{ \tilde{\rho}(C/C_0 = c_m) }{ \tilde{\rho}_1 }$. In figure \ref{fig:nonmonotonic_densities}(a), we plot the spatial variation of the non-monotonic density profile defined in equation \eqref{eq:nonmonotonic3} for $\rho_m = 1.8, \; t_0 = 25$, and Ra = 200 along with the base-state concentration, $c_b(x)$ in figure \ref{fig:nonmonotonic_densities}(b). As seen from figure \ref{fig:nonmonotonic_densities}(a), our model non-monotonic density profiles are qualitatively similar to the true density profiles obtained in various reaction-diffusion and double diffusive systems \cite{Trevelyan2011, Lemaigre2013, Loodts2015, Loodts2016}. 

\section{Results and discussion \label{sec:results}}

\begin{figure}[hbtp]
\centering
\includegraphics[scale=0.5]{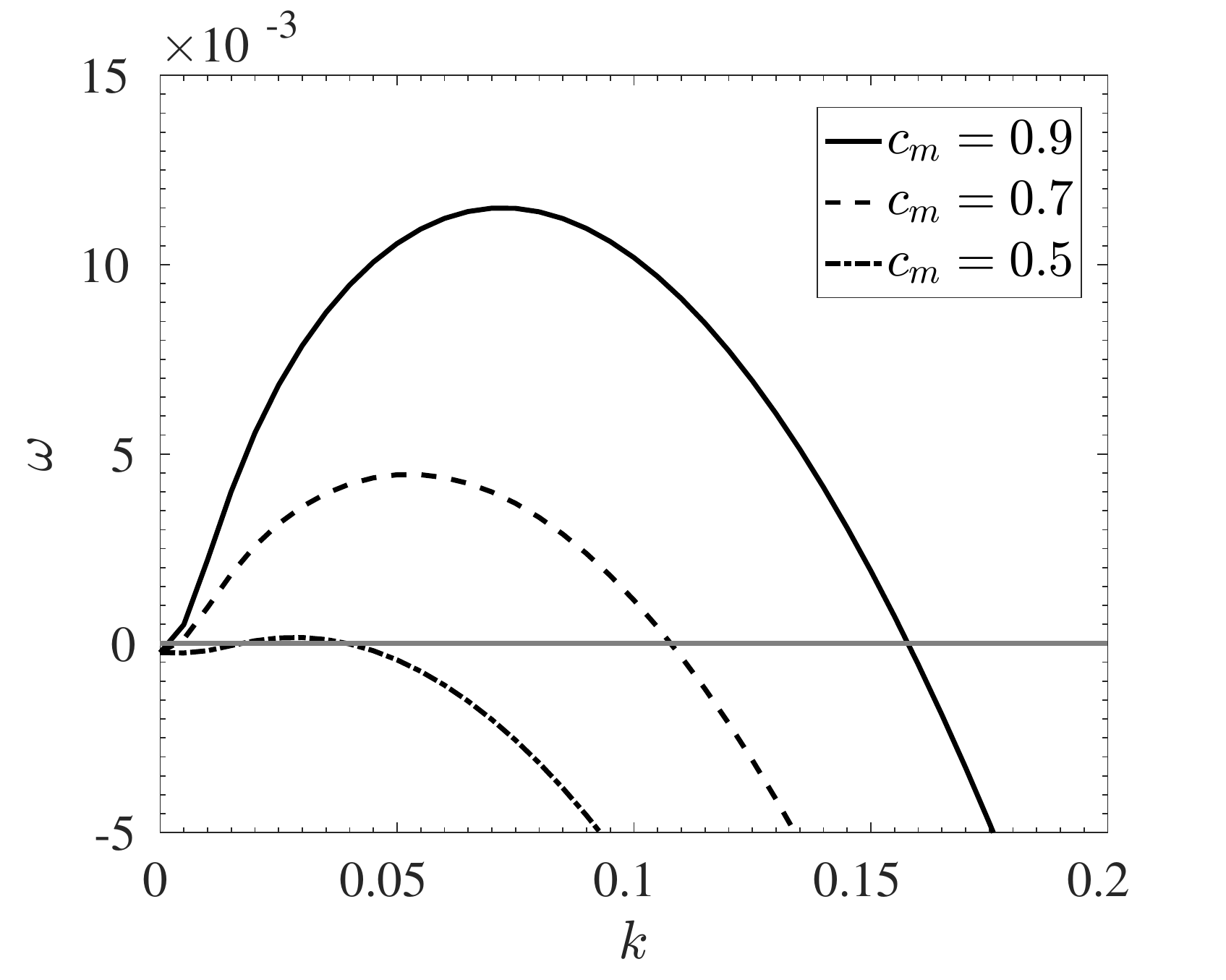} 
\caption{Dispersion curves obtained from the linear theory corresponding to the density profiles shown in figure \ref{fig:nonmonotonic_densities}(a).} 
\label{fig:nonmonotonic_dispersion_curves}
\end{figure}

\subsection{Dispersion curves \label{subsec:dispersion_nonmonotonic}} 

\begin{figure*}[hbtp]
\centering
(a) \\ 
\includegraphics[scale=0.6]{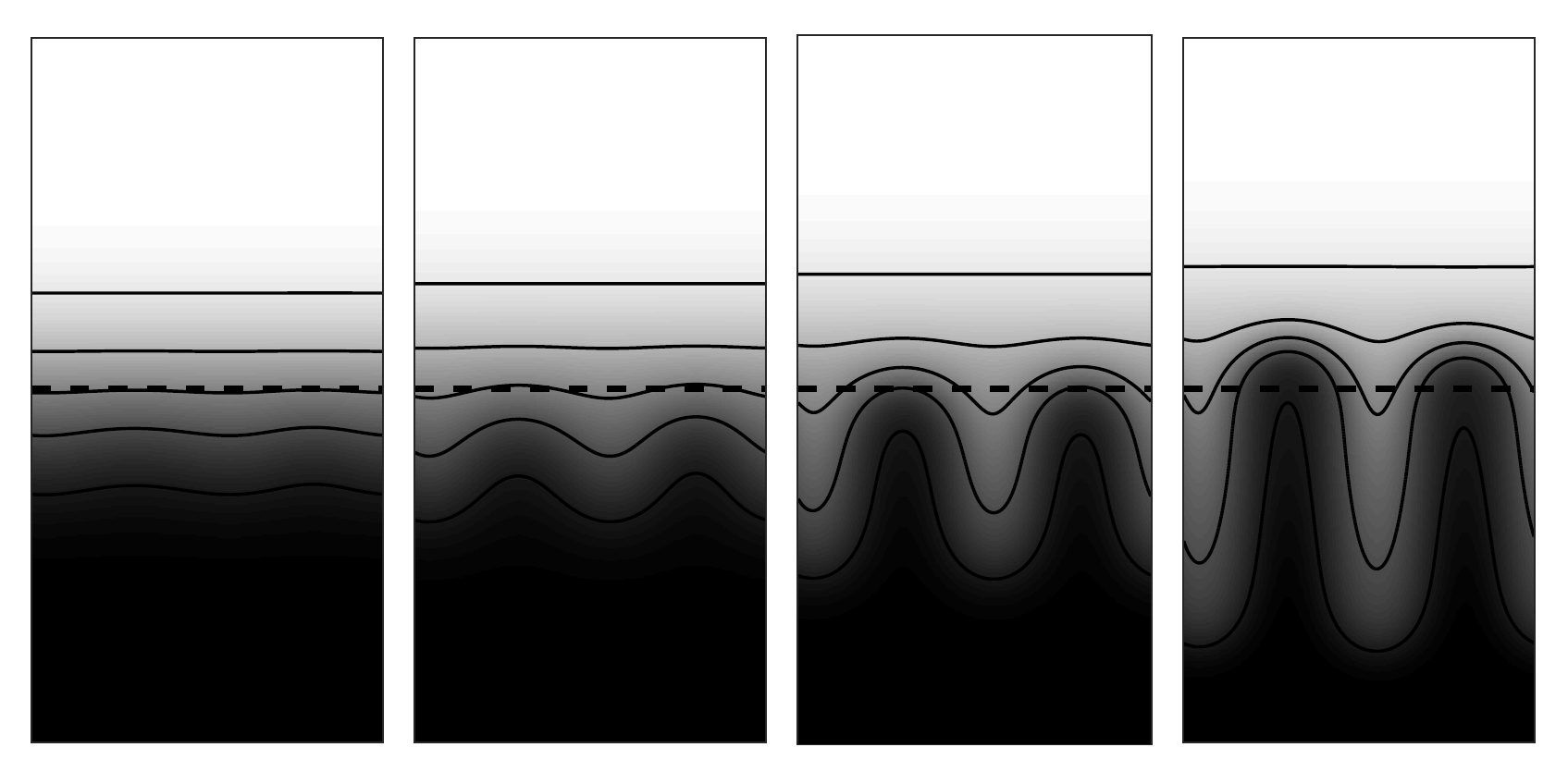} \\ 
(b) \\ 
\includegraphics[scale=0.6]{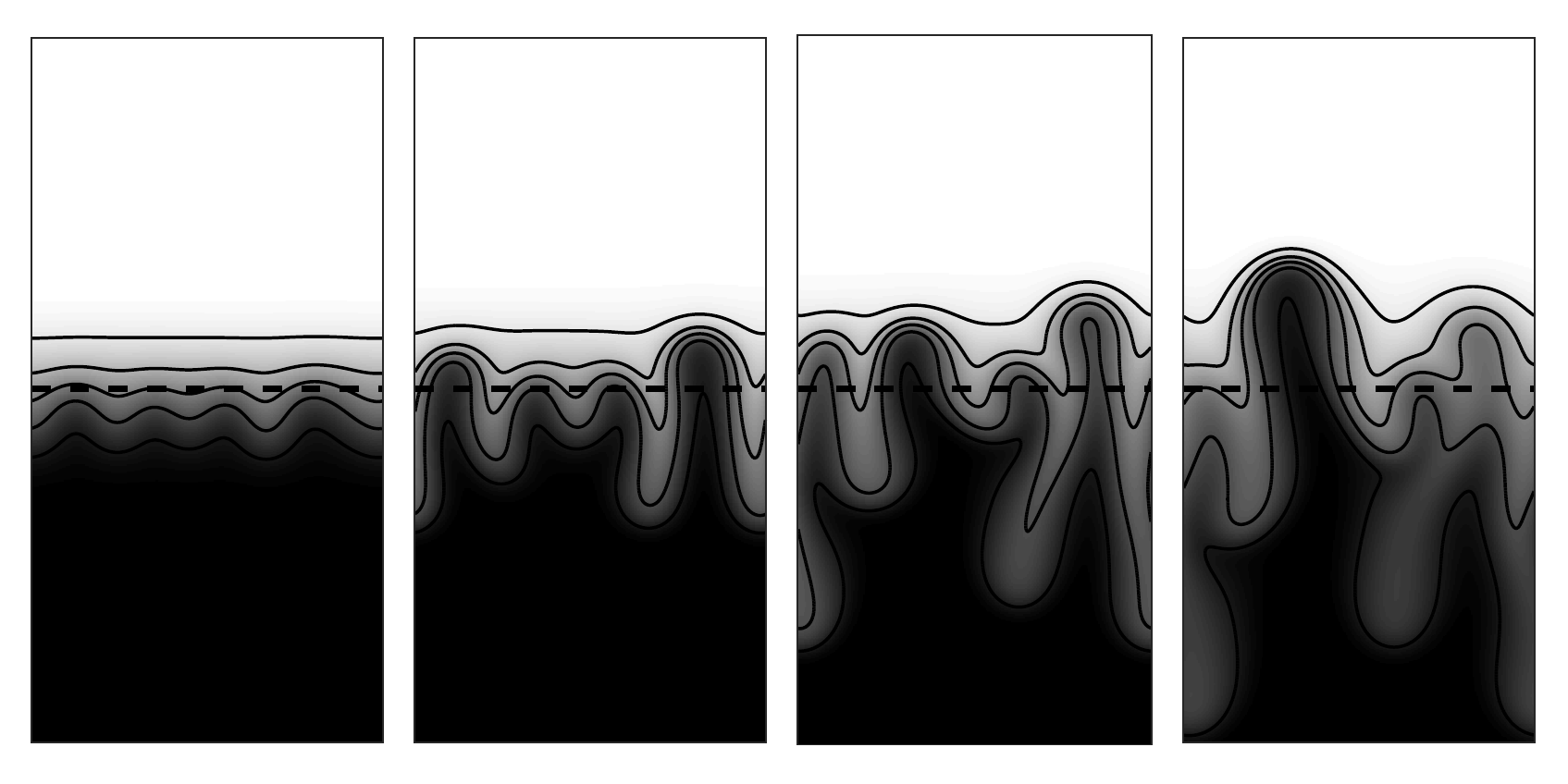} \\ 
(c) \\ 
\includegraphics[scale=0.6]{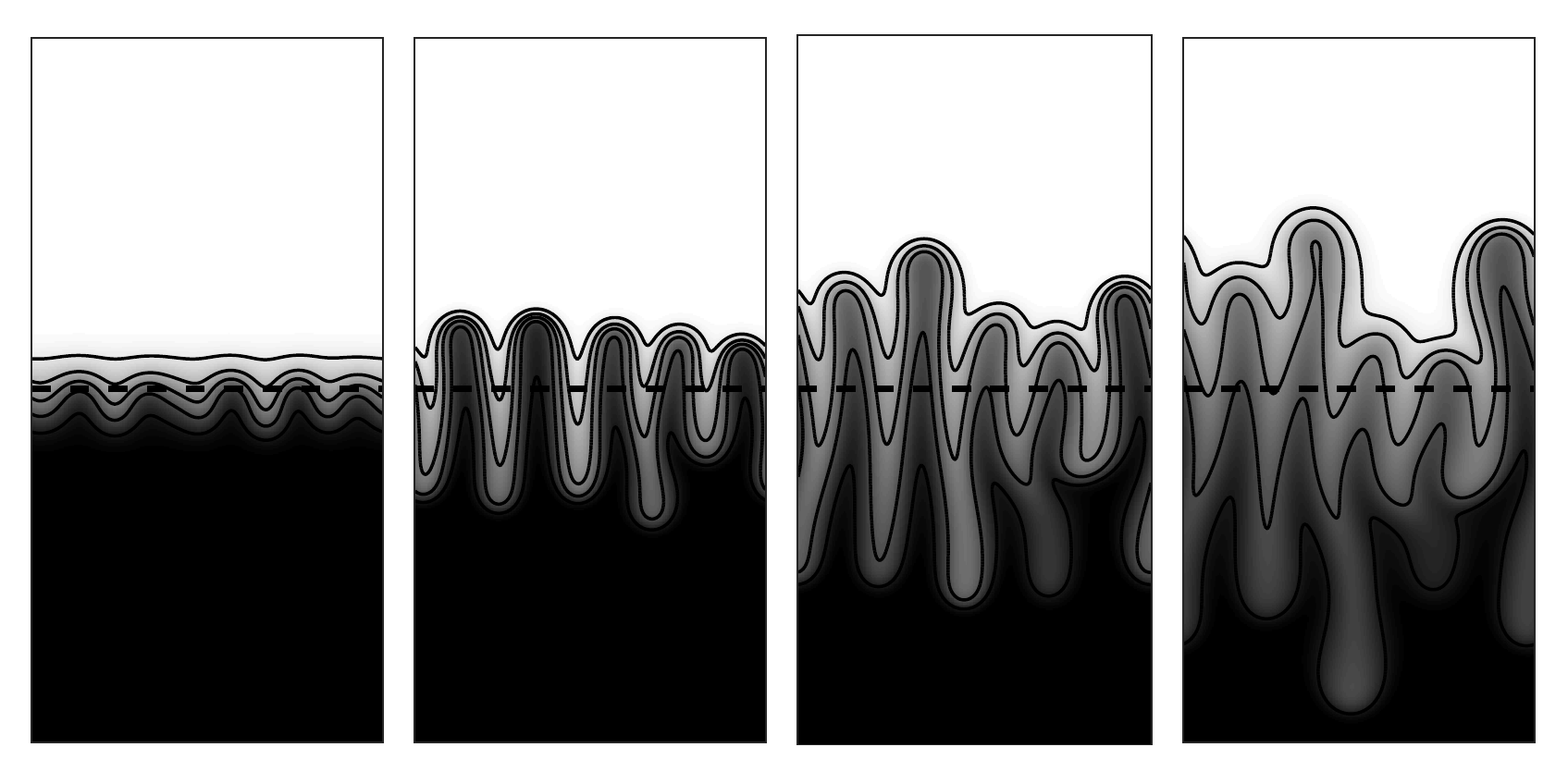}
\caption{Asymmetric Rayleigh-Taylor fingers for the non-monotonic density profile $\rho_{{\rm nm}, 3}$ for Ra = 1024: (a) $c_m = 0.5$; (From left to right $t = 25000, \; 30000, \; 35000, \; 40000$.) (b) $c_m = 0.7$; (From left to right $t = 7500, \; 10000, \; 12500, \; 14500$.) (c) $c_m = 0.9$. (From left to right $t = 3000, \; 4500, \; 6000, \; 7500$.) The horizontal dashed line represents the position of the initial interface.} 
\label{fig:NLS_nonmonotonic} 
\end{figure*}

Linear stability analysis is performed for the density profiles shown in figure \ref{fig:nonmonotonic_densities}(a) and the corresponding dispersion curves are shown in figure \ref{fig:nonmonotonic_dispersion_curves}. 
For the non-monotonic density profiles $\tilde{\rho}(c)$, $\rho_m$ (alternatively, $\Delta \rho = \rho_m - 1$) remains unchanged, but the corresponding value of the concentration $c_m$ changes that in turn changes the density gradient in both the stable and unstable zones. As $c_m$ decreases, the location of the density maximum appears more towards the initial interface ($x = 0$)--the unstable density stratification reduces, while the stable density stratification increases. Note that the density profiles obtained corresponding to $c_m = 1$ ($c_m = 0.5)$ is monotonic (symmetric about $x = 0$). Whereas, reducing $c_m$ from $1^{-}$ through $0.5^{+}$ we obtain a family of asymmetric non-monotonic density profiles. Here, we choose $c_m = 0.9, 0.7$ and $0.5$. The corresponding dispersion curves shown in figure \ref{fig:nonmonotonic_dispersion_curves}(c) depict that the least unstable system is obtained for $c_m = 0.5$--the symmetric density profile. The most unstable wavenumber and the corresponding growth rate increases as $c_m$ increases. In response, in the nonlinear regime we anticipate long wave, rounded convective fingers to form at a later time for $c_m = 0.5$, whereas short wave, sharp convective fingers for $c_m = 0.9$, and it is evident from our numerical simulation results shown in figure \ref{fig:NLS_nonmonotonic}. The qualitative response of the dispersion curves to the variation of the density gradients are in accordance to that discussed above. By parity of reasoning we can argue that the instability is further delayed as $c_m$ is further decreased within the interval $c_m \in (0, 0.5)$, most probably a linearly stable state is obtained as $c_m \rightarrow 0^{+}$. It is noteworthy that $c_m < 1/2$ imposes the following restriction on the choice of $\rho_m$: 

\begin{equation}
\label{eq:rhom_restriction} 
\rho_m < 1 + \frac{ c_m^2 }{ 1 - 2 c_m }. 
\end{equation}
This is essential to insure that the dimensional density is always positive and physically realistic. 

\begin{figure*}[hbtp]
\centering
(a) \hspace{3.2 in} (b) \\ 
\includegraphics[scale=0.5]{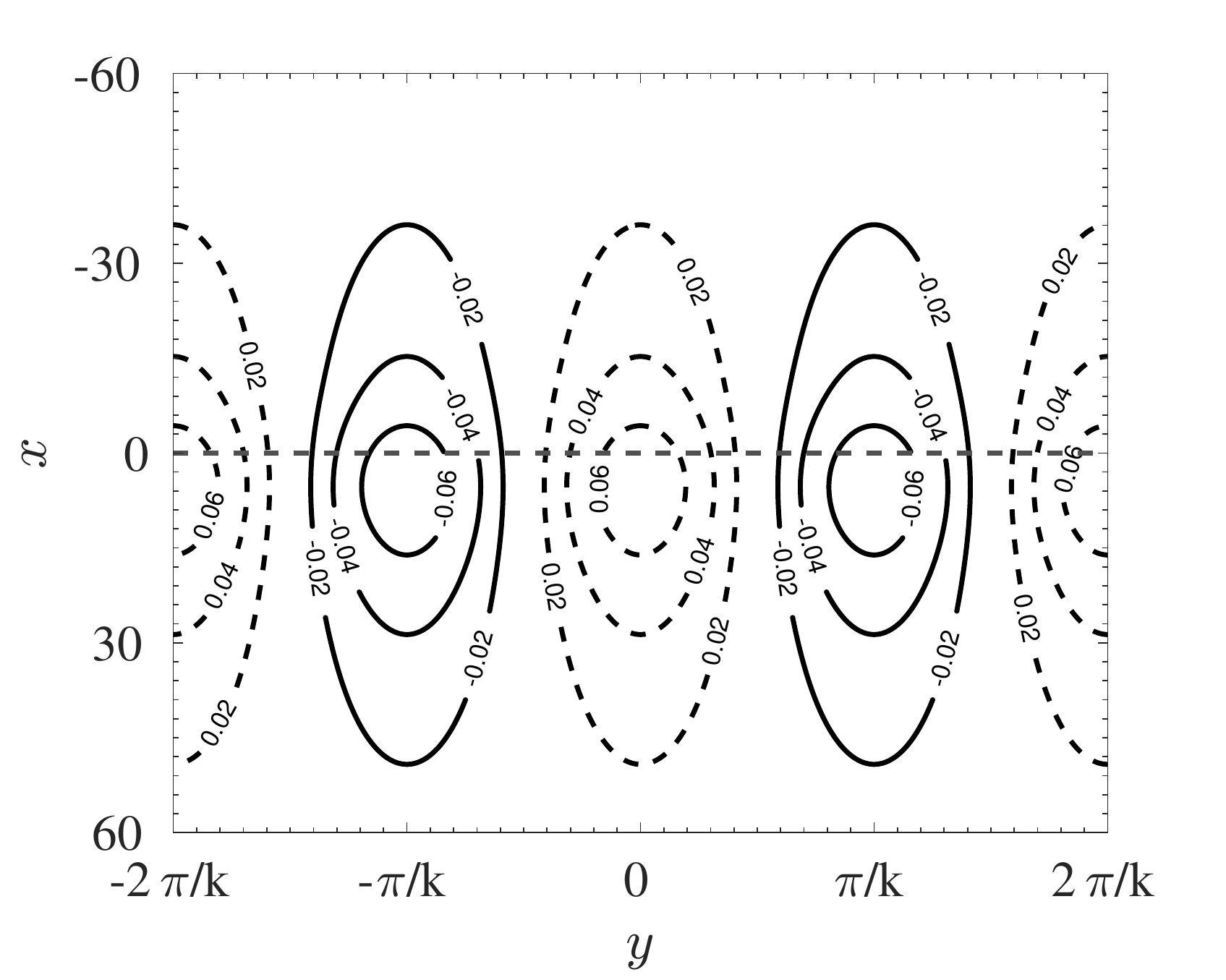} 
\includegraphics[scale=0.5]{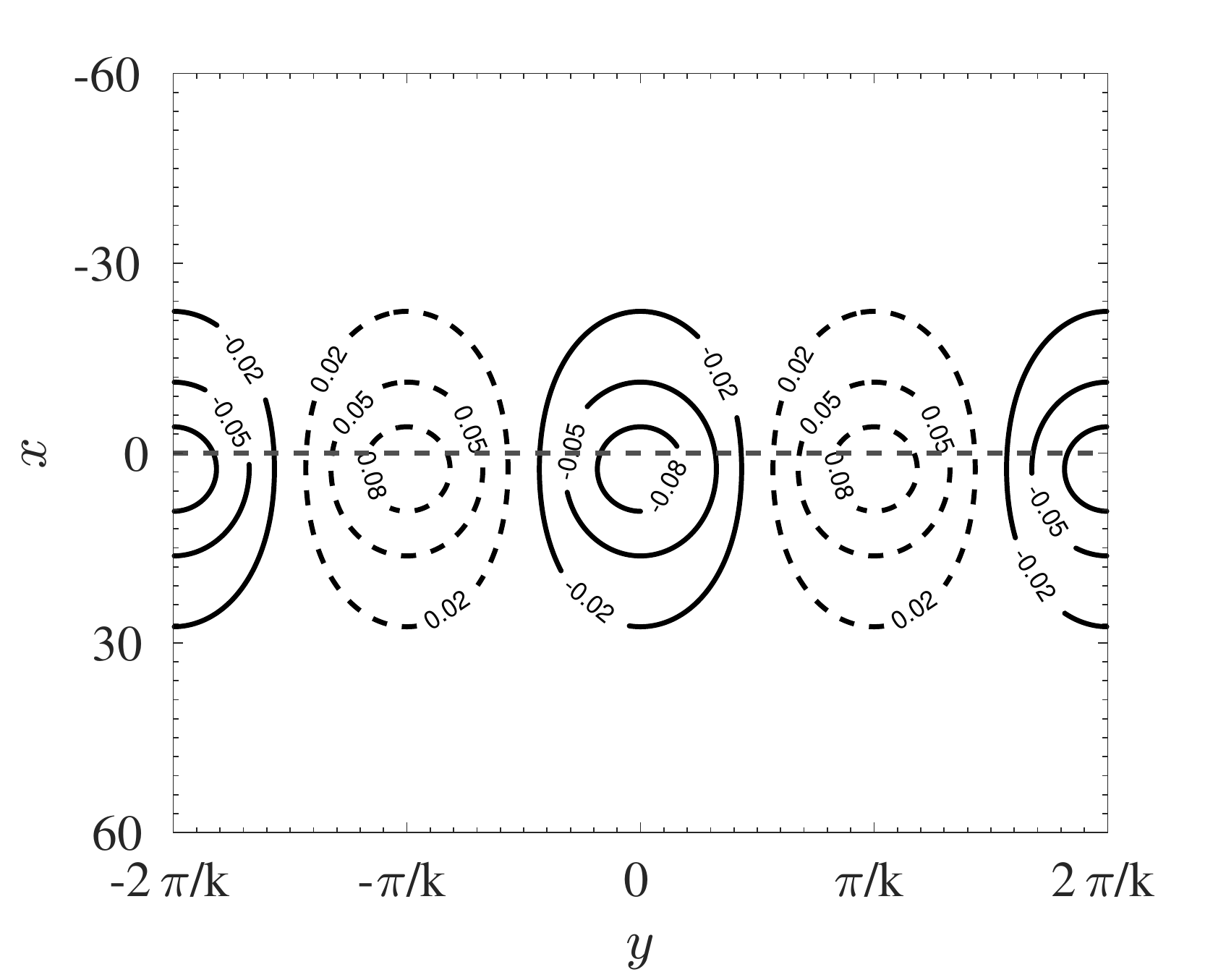}  \\ 
(c) \\ 
\includegraphics[scale=0.5]{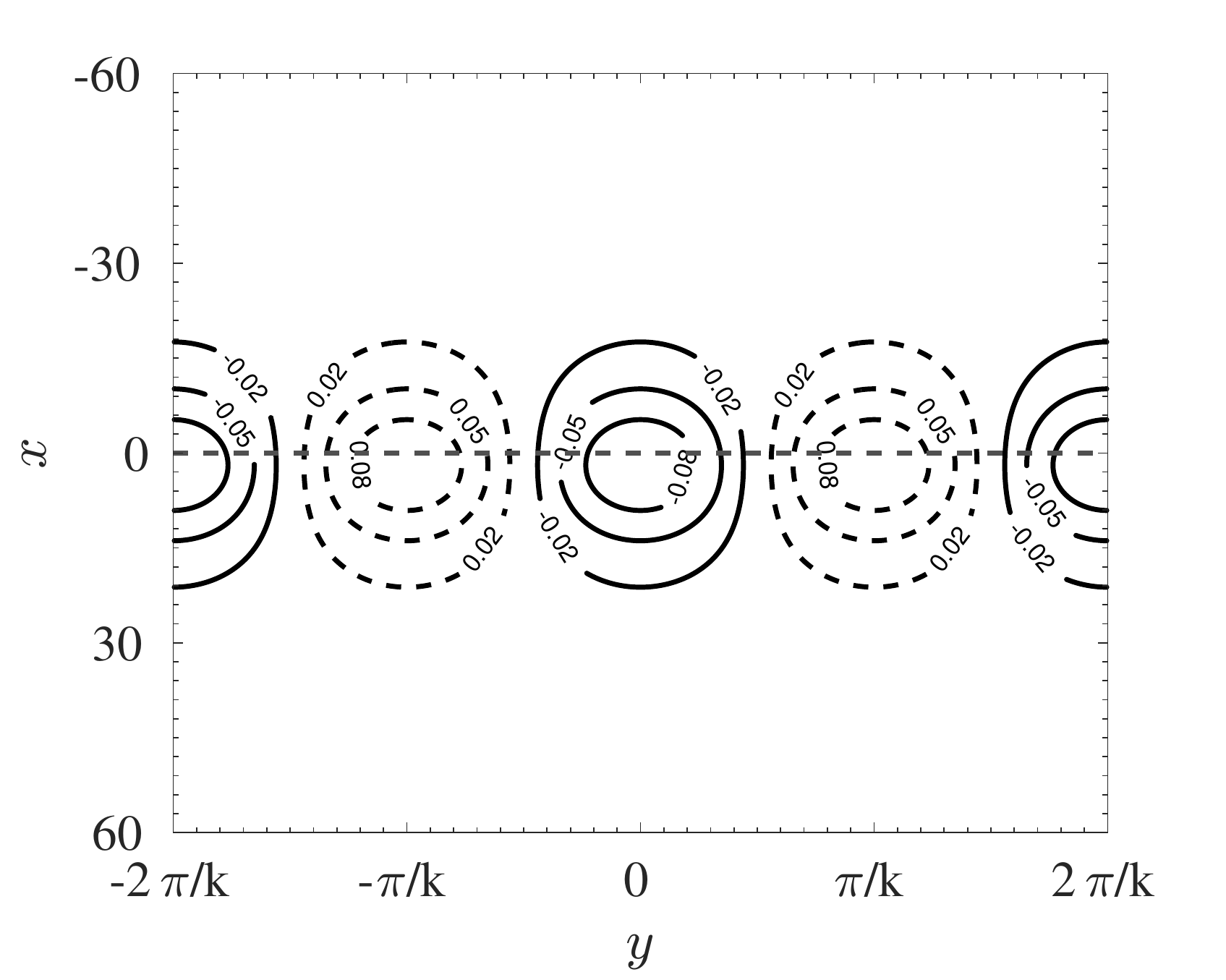}
\caption{The most unstable eigenfunction obtained from QSSA for $t_0 = 25$, and Ra = 200. (a) $c_m = 0.5, \; k = 0.03$; (b) $c_m = 0.7, \; k = 0.05$; (c) $c_m = 0.9, \; k = 0.07$. The horizontal dashed line represents the position of the initial interface. 
} 
\label{fig:Evec_nonmonotonic} 
\end{figure*}

\subsection{Eigenfunctions \label{subsec:eigenfunctin_nonmonotonic}} 

In figure \ref{fig:Evec_nonmonotonic}, we plot the contours of the eigenfunction corresponding to the most unstable wavenumber estimated from the dispersion curves shown in figure \ref{fig:nonmonotonic_dispersion_curves}. As the unstable density gradient increases (i.e., $c_m$ increases from 0.5 through 0.9 in figures \ref{fig:Evec_nonmonotonic}(a) through \ref{fig:Evec_nonmonotonic}(c)), the eigenfunctions are more localized near the initial position of the interface $x = 0$. Moreover, as the stable density gradient increases and the unstable density gradient decreases, the eigenfunction becomes more asymmetric about the initial interface. 

\subsection{Nonlinear simulations \label{subsec:NLS_nonmonotonic}} 

Our linear stability results are supported with nonlinear simulations. When the density maximum occurs at the interface between two fluids, the density profile is symmetric about the interface--the density gradient at the same distance on either side of the interface is of the same magnitude, but of different sign. Although the stable density gradient is of equal strength to that of the unstable density gradient, it does not completely suppress the instability. Indeed, one anticipates long wave finger formation at a later time. Physically speaking, as the unstable density gradient moves towards a smaller value--the density maximum occurs in the defending fluid--convective motion sets in within the diffusive boundary below the density maximum and convection pushes the boundary of the density maximum slightly upwards against the resistance from the stable density gradient. This in turn reduces the strength of the stable density gradient and hence the fingers penetrate into the displacing fluid. Figure \ref{fig:NLS_nonmonotonic} shows the concentration maps for $c_m = 0.5$, $0.7$ and $0.9$, which depict the behavior explained above. 

\begin{figure}[hbtp]
\centering
\includegraphics[scale=0.5]{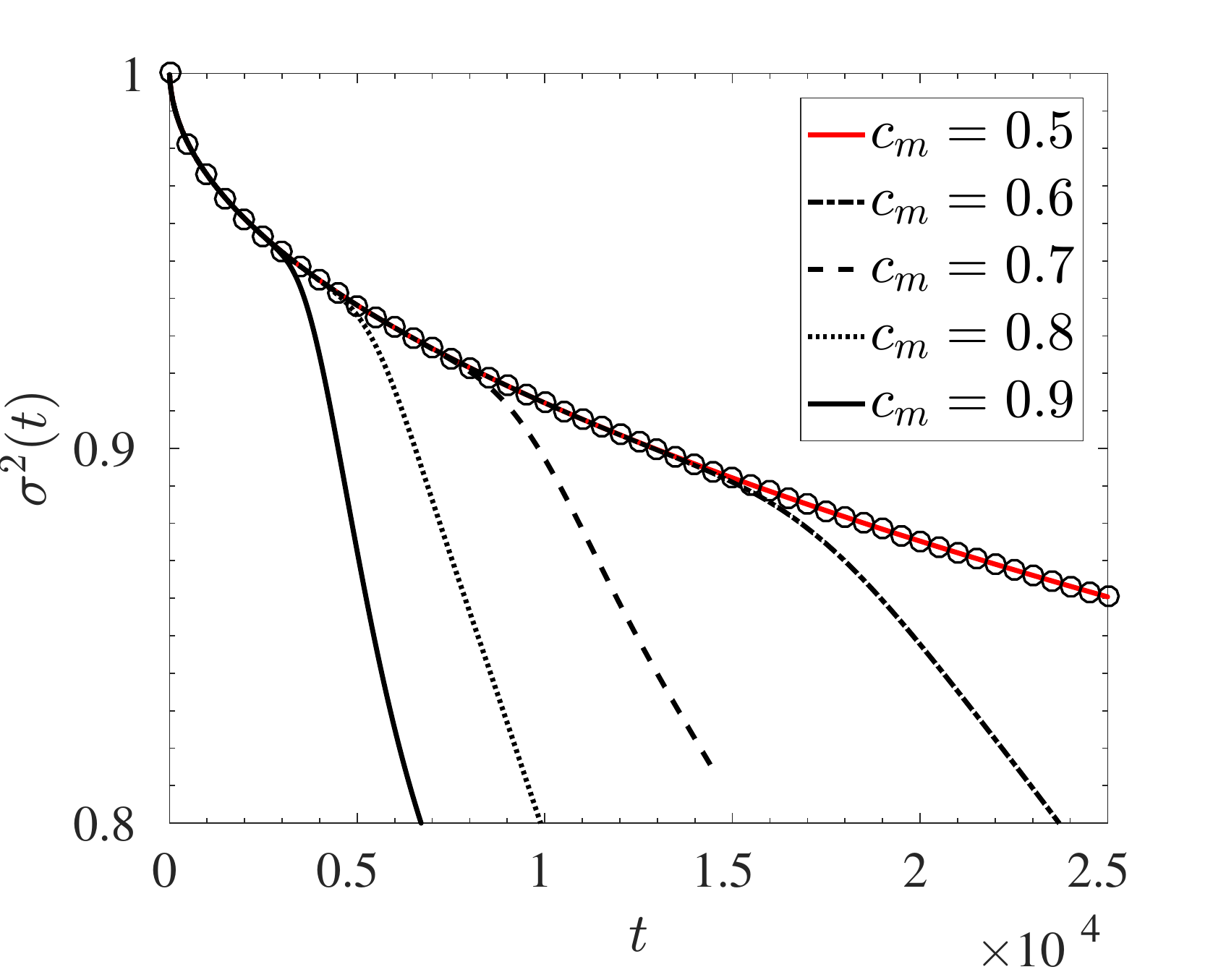} 
\caption{(Color online) Temporal evolution of the variance, $\sigma^2$, for Ra = 1024 for $c_m = 0.5, \; 0.6, \; 0.7, \; 0.8$ and $0.9$. Circles correspond to the stable flow in the absence of any density stratification.} 
\label{fig:var}
\end{figure}

Therefore, the Rayleigh-Taylor instability in a miscible porous media flow is better described when Ra is combined with the density gradient, $ \displaystyle \frac{\partial \rho}{\partial x} = \rho^{\prime}(c) \frac{ \partial c }{ \partial x } $, alternatively in terms of Ra$_g$ as defined in equation \eqref{eq:Rayleigh_new}. Since $\rho^{\prime}(c)$ is fixed for a given system, from the linear theory we conclude that instability depends on $\displaystyle \frac{ \partial c }{ \partial x } $, which varies continuously with $t_0$. It is observed that an increasing magnitude of the stable density gradient reduces the growth rate. As a resultant, the fluid mixing is mitigated, which we quantify in terms of the variance \cite{Pramanik2015c} 

\begin{equation}
\label{eq:variance}
\sigma^2(t) = \langle c^2 \rangle - \langle c \rangle^2, 
\end{equation}
where $\langle \cdot \rangle$ represents an average over the spatial domain. Figure \ref{fig:var} depicts that the variance decreases faster when the density maximum appears in the penetrating fluid, representing an enhanced mixing between the defending and penetrating fluids as the convective instability increases. The temporal variation of $\langle \sigma^2 \rangle$ corresponding to a buoyantly unstable case deviates from that of a stable flow when the convection starts. This demarcates the onset of instability. It is observed from figure \ref{fig:var} that for $c_m = 0.5$, the convective instability is yet to set in at $t \approx 2.5 \times 10^4$, wherein convection starts as early as $t \approx 5 \times 10^3$ for $c_m = 0.9$. We verified that when density varies linearly in $c$, instability sets in at an earlier time than the non-monotonic profiles defined in equation \eqref{eq:nonmonotonic3}. 

\section{Discussion and conclusions \label{sec:discuss}}  

Motivated by the basic questions in porous media convection, we developed a phenomenological theory to treat density gradients in the context of the dynamics of convective instability in porous media flows. The principal phenomenon capturing our interest is the influence of the local density gradients on the convective instability. By parity of reasoning as the location of the density maximum appears above (below) the fluid interface the instability enhances (diminishes) as compared to the situation when the density maximum appears at the fluid interface. In common practice, the stability characteristics of the miscible RT fingers in porous media are described in terms of the dimensionless group Rayleigh-Darcy number (Ra) defined in equation \eqref{eq:Rayleigh}. For a steady-state base flow, convection starts for Ra above a critical value, Ra $>$ Ra$_c$. For a time-dependent base flow, the concept of critical Rayleigh-Darcy number is replaced by the critical time of instability (onset of instability) $t_c$ \cite{Riaz2006}. Surprisingly, for the same Ra, one obtains different $t_c$ depending on the density-concentration relation, which may lead to different interpretations of the miscible RT fingers. 

Here, we present a unified theory in terms of the density gradients. Through a systematic analysis of a series of density profiles, we are able to identify the effects of density gradients on miscible Rayleigh-Taylor fingering. Our density profiles can be uniquely defined as 

\begin{equation}
\label{eq:density_general} 
\tilde{\rho}(c) = \tilde{\rho}_1 + \Delta \rho f(c), 
\end{equation}
wherein 

\begin{eqnarray}
\label{eq:fc_deltarho_nonmonotonic3}
f(c) = -c(c-2c_m), \;\;\; \Delta \rho = \tilde{\rho}_1 \frac{(\rho_m - 1)}{c_m^2}, 
\end{eqnarray}
for the non-monotonic density profiles \eqref{eq:nonmonotonic3}. The competition between a locally stable and a locally unstable density gradients are discussed through various non-monotonic profiles. Our model construction allows to single out the effects of variation in stable as well as unstable density gradients. Interestingly, the explanation of miscible Rayleigh-Taylor fingers through the density gradients presented here has important implications for geological CO$_2$ sequestration and other porous media convection driven out of chemical reactions \cite[and refs. therein]{DeWit2016, Loodts2016}, where the fluid density varies non-monotonically on the solute concentration. One of the most important results of our theory is the following. The role of density gradients can give a possible explanation of the asymmetric RT finger formation in double diffusive reactive system \cite{Lemaigre2013} without solving the multi-component reaction-diffusion-convection equations.  

As shown in figure \ref{fig:nonmonotonic_densities}(a), the unstable density gradient, $ \displaystyle \left[ \frac{ \partial \rho }{ \partial x } \right]_{x > x_m} = \left[ f^{\prime}(c) \frac{ \partial c }{ \partial x } \right]_{x > x_m}$, decreases as the density maximum appears higher above the initial fluid interface, which we mark at $c = 0.5$ (i.e., $x = 0$, which is obtained by parity of symmetry of the error function profile). Here, $x_m$ is obtained inverting the relation $c_m = c(x_m, t_0)$, i.e., $x_m$ depends on $t_0$. Additionally, the magnitude of the stable density gradient, $ \displaystyle \left[ \frac{ \partial \rho }{ \partial x } \right]_{x < x_m} = \left[ f^{\prime}(c) \frac{ \partial c }{ \partial x } \right]_{x < x_m}$, increases as $c_m$ decreases. Note that although Ra = 200 remains unchanged for the three density profiles considered (figure \ref{fig:nonmonotonic_densities}), the corresponding dispersion curves are different (figure \ref{fig:nonmonotonic_dispersion_curves}). This signifies that if the density profile of a physical system evolves non-monotonically having different local density gradients without affecting Ra, one may over/under predict the stability. Therefore, an \emph{a priori} estimate about the stability of a system for a given Ra may be misleading. Contrary to that, a quantification of the stability of a system in terms of the local density gradient as discussed above improves our understanding. This confirms that when the solute concentration field possesses transient nature, the Rayleigh-Taylor instability in porous media is better explained in terms of the local density gradients, but not the difference between the maximum and minimum density. 

Recently, Toppaladoddi and Wettlaufer \cite{Srikanth2017} studied high Rayleigh numbers penetrative convection of a fluid confined between two plates. One of the major results of their work was to identify the essence of a dimensionless parameter ($\Lambda$)--the ratio of the top and bottom plate temperatures relative to the temperature that maximizes the water density--apart from Rayleigh and Prandtl numbers. In particular, it was reported that the the penetration of the plumes generated from the hot bottom plate is hindered by a strong stable layer stratification for a large $\Lambda$. Similarly, our results corresponding to the density profile $\tilde{\rho}_{{\rm nm}, 3}$ can be explained in terms of the dimensionless parameter 
\begin{equation}
\label{eq:Gamma}
\Gamma = \frac{1-c_m}{c_m} > 0. 
\end{equation}
It is clear from figures \ref{fig:NLS_nonmonotonic}(a-c), which depict time evolution of the concentration field for $\Gamma = 1, 3/7, 1/9$, respectively, that due to strong stable layer stratifications the fingers are impeded to penetrate into the stable layer as $\Gamma$ increases, and the flow remains stable for a longer period. As $\Gamma$ decreases, the fingers penetrate the stable upper layer and the patterns approach toward an up-down symmetric situation. Therefore, it can be expected that the standard Rayleigh-Taylor fingers with a linear density-concentration relation is neared as $\Gamma \rightarrow 0$. 

However, we must note that although the qualitative dependences were understood in the course of this study, the quantitative relation between the growth rate and the density gradients remained unclear. Note that most of the recent studies on stability analysis of miscible fingering instabilities have emphasized on various other stability methods--QSSA with respect to a suitably defined self-similar variable, $\xi$ (SS-QSSA) \cite{Pramanik2013}, initial value calculation (IVC) \cite{Hota2015a}, non-modal stability analysis (NMA) \cite{Hota2015b}, etc.--than the classical QSSA. Recall, the instantaneous growth rates obtained using SS-QSSA and QSSA are related via \cite{Kim2011}

\begin{equation}
\label{eq:growth_rate_relation}
\omega_{x} = \omega_{\xi} + \frac{ 1 }{ 4 t_0 }, 
\end{equation}
where $t_0$ is the frozen diffusive time \cite{Pramanik2013}. The present study focuses upon the role of local density gradients on Rayleigh-Taylor fingers in a homogeneous porous medium. A qualitative understanding without any direct comparison of with experiments allows to choose QSSA method for the stability analysis. We believe the qualitative results presented here remains unaffected by the choice of the method. Interested researchers are encouraged to choose the most appropriate stability method for a more quantitative measurements, in particular while comparing with the experiments. 

We show that the onset of instability, and hence the fluid mixing can be significantly altered by varying the stable and unstable density gradients parameterized in terms of $\Gamma$. This indicates that chemical reactions, that result a continuous evolution of the density gradients, control convective instability and fluid mixing. Mathematically, convective instability occur for any density profile as long as density decreases along the direction of gravity--the onset of instability depends on the gradients of the density in the stable as well as unstable zones. Diffusion causes to change the density gradient in both the stable and unstable zones. Convection sets in early for a steep (shallow) unstable (stable) density gradient and the onset time  increases as the unstable (stable) density gradient becomes shallow (steep). Present analysis can be extended to understand miscible viscous fingering with non-monotonic viscosity profiles \cite{Manickam1993, Haudin2016}. A more complicated situations of non-monotonic density effects on the convective instability in porous media awaits understanding when the solute concentration is localized within a finite region \cite{DeWit2005, Mishra2008, Pramanik2015e}. 

\begin{acknowledgements}
SP thanks Prof. John Wettlaufer, Cristobal Arratia, and Francesco Macarella for many helpful discussions. This work started when SP was at the Department of Mathematics, IIT Ropar and was generously funded a Ph.D. fellowship by the National Board for Higher Mathematics (NBHM), Department of Atomic Energy (DAE), Govt. of India. 
\end{acknowledgements}


%





\end{document}